\documentclass[twocolumn,twoside,slac_two]{revtex4}
\usepackage{graphicx}
\usepackage{fancyhdr}
\newcommand{\ana}[1]{\textit{Astronomy \& Astrophysics},}
\newcommand{\aap}[1]{\textit{Astronomy \& Astrophysics},}
\newcommand{\aas}[1]{\textit{Astronomy \& Astrophysics Supplement Series},}
\newcommand{\aaps}[1]{\textit{Astronomy \& Astrophysics Supplement Series},}
\newcommand{\acta}[1]{AcA,}
\newcommand{\aj}[1]{\textit{Astronomical Journal},}
\newcommand{\annrev}[1]{\textit{Annual Reviews of Astronomy \& Astrophysics},}
\newcommand{\apj}[1]{\textit{Astrophysical Journal},}
\newcommand{\apjl}[1]{\textit{Astrophysical Journal Letters},}
\newcommand{\apjs}[1]{\textit{Astrophysical Journal Supplement Series},}
\newcommand{\araa}[1]{\textit{Annual Reviews of Astronomy \& Astrophysics},}
\newcommand{\baas}[1]{\textit{Bulletin of the American Astronomical Society},}
\newcommand{\mnras}[1]{\textit{Monthly Notices of the Royal Astronomical Society},}
\newcommand{\pasp}[1]{\textit{Publications of the Astronomical Society of the Pacific},}

\begin{document}

\title{Multiwavelength Observations of Gamma-ray Binary Candidates}

%

\author{M.\ Virginia McSwain}
\affiliation{Lehigh University, 16 Memorial Drive East, Bethlehem, PA 18015, USA}
\author{Masha Chernyakova}
\affiliation{Dublin City University, Dublin, Ireland}

\author{Denis Malishev}
\affiliation{Bogolyubov Institute for Theoretical Physics, Kiev, Ukraine}

\author{Michael De Becker}
\affiliation{Universit\'e de Li\`ege, Address, Li\`ege, Belgium}

\author{Stephen Williams}
\affiliation{Georgia State University, Atlanta, GA, USA}

\begin{abstract}
A rare group of high mass X-ray binaries (HMXBs) are known that also exhibit MeV, GeV, and/or TeV emission (``gamma-ray binaries'').  Expanding the sample of gamma-ray binaries and identifying unknown Fermi sources are currently of great interest to the community.  Based upon their positional coincidence with the unidentified Fermi sources 1FGL J1127.7--6244c and 1FGL J1808.5--1954c, the Be stars HD 99771 and HD 165783 have been proposed as gamma-ray binary candidates.  During Fermi Cycle 4, we have performed multiwavelength observations of these sources using \textit{XMM-Newton} and the CTIO 1.5m telescope.  We do not confirm high energy emission from the Be stars.  Here we examine other X-ray sources in the field of view that are potential counterparts to the Fermi sources.
\end{abstract}

\maketitle

\thispagestyle{fancy}


\section{Introduction}

There are a small but growing number of high mass X-ray binaries (HMXBs) that also exhibit emission above 100 MeV.  Currently, this category of ``$\gamma$-ray binaries'' includes:  
LS I +61 303 \citep{albert2006, abdo2009a}, 
LS 5039 \citep{aharonian2006,  abdo2009b}, 
Cygnus X-3 \citep{abdo2009c}, 
1FGL  J1018.6--5856 \citep{ackermann2012}, 
PSR B1259--63 \citep{aharonian2005, abdo2011}, 
HESS J0632+057 \citep{aharonian2007, hinton2009},  
and possibly Cygnus X-1 \citep{albert2007, sabatini2013}.  

Each of the $\gamma$-ray binaries contain a hot, young star (either Wolf-Rayet, O, or Be spectral class) and exhibit variable emission across the electromagnetic spectrum.  In the first four systems, the MeV--GeV light curve is modulated with the orbital period determined from optical spectra and/or radio flux measurements.  Furthermore, all three were identified in the \textit{Fermi} Year 1 Point Source Catalog (1FGL) using the binary orbital period as a key signature \citep{abdo2010}.  The MeV--TeV emission of PSR B1259--63 also correlates with the source orbital phase in that it has only been successfully detected close to the periastron phases (\citealt{aharonian2005}; \citealt{abdo2011}).  
The high energy emission of HESS J0632+057 is also variable with a period of 321 days, similar to the proposed optical counterpart (\citealt{bongiorno2011}; \citealt{casares2012}).

About 60\% of HMXBs are Be/X-ray binaries (BeXRBs) that contain a B-type star with a circumstellar mass-loss disk (a Be star) with a compact companion, usually a neutron star \citep{liu2006}.  Be star systems also comprise about 50\% of the known gamma-ray binaries:  LS I +61 303, PSR B1259--63, and HESS J0632+057.  The high energy emission is powered by either wind accretion onto the compact star, or by the collision of stellar and relativistic pulsar winds in a shock region \citep{dubus2006}.  All of these sources present a unique opportunity to study particle acceleration in nearby, Galactic sources.  

We have cross-correlated the 1FGL \citet{abdo2010} with the \citet{jaschek1982} catalog of known Be stars, and we discovered two Be stars within the 95\% error ellipse of 1FGL sources.  Using \textit{XMM-Newton} images, we find that the Be stars are not significant X-ray sources and are unlikely counterparts of the very high energy emission.

\section{Observations of 1FGL J1808.5--1954c (2FGL 1808.6--1950c)}

\begin{figure}
\includegraphics[width=75mm]{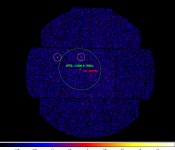}
\caption{A 17.8 ks XMM-Newton MOS1 image of the HD 168783 field of view.  The GeV source position and 95\% error ellipse are shown in green. The position of the Be star is marked as a red cross, but no keV emission is detected at that position. However, two unidentified keV sources lie within or near the 2FGL error ellipse. }
\label{mos1}
\end{figure}

The GeV source position is known to within 4.7 arcmin (95\% confidence) and may be associated with the Be star HD 165783.  The optical star has been correlated with a serendipitous \textit{XMM-Newton} source \citep{watson2009}, but the pipeline images suffer from very high background and the source lies near the chip gaps.  Therefore we obtained new \textit{XMM-Newton} observations of the field on 2012 March 24, shown in Figure \ref{mos1}.  No keV emission is detected at the position of the proposed optical counterpart HD 165783.  

\begin{figure}
\hspace{-0.55in}
\includegraphics[angle=180, width=95mm]{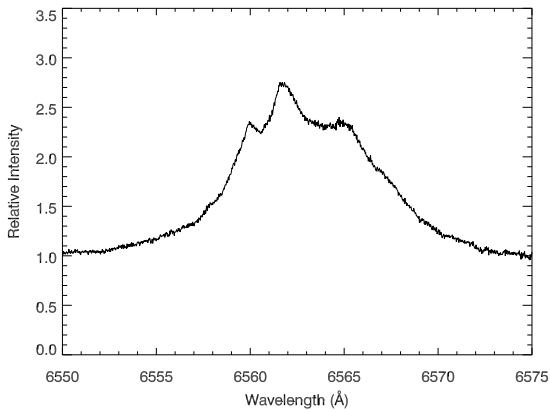}
\vspace{-0.3in}
\caption{The H$\alpha$ line profile of the Be star HD 168783 reveals a complex circumstellar disk structure that makes it difficult to determine the starÕs radial velocity. }
\label{halpha}
\end{figure}

We observed HD 165783 using the CTIO 1.5m telescope, operated by the SMARTS consortium, and CHIRON spectrograph between 2011 August and 2011 November.  The star has a spectral type of B3/5 Ve.  It has a high proper motion of 6.4 $\mu$as yr$^{-1}$.  We are unable to determine its radial velocity due to the very complex structure of the H$\alpha$ emission, shown in Figure \ref{halpha}, so its runaway nature cannot be determined.  No orbital period is known.  This star may be a post-supernova runaway system.

The counts map of 1FGL J1808.5--1954c reveals this source within the crowded Galactic plane region.  Since the time our Fermi Cycle 4 proposal was written, the GeV source has been associated with a globular cluster \citep{nolan2012}.

\begin{figure}
\includegraphics[width=75mm]{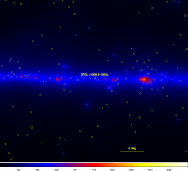}
\caption{Fermi counts map of 1FGL J1808.5--1954c.  Error ellipses of nearby sources from the 2FGL are shown in yellow.  }
\label{cmap}
\end{figure}

 \begin{figure}
 \includegraphics[width=75mm]{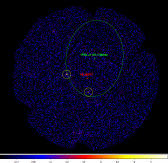}
 \caption{A 13.5 ks \textit{XMM-Newton} MOS2 image of the HD 99771 field of view.  The GeV source and Be star positions are marked in the format of Figure 1.  Once again, no keV emission is detected at that the Be starÕs position. Two unidentified keV sources lie within or near the 1FGL error ellipse.  \label{mos2}}
 \end{figure}

\clearpage
 \begin{figure}
 \hspace{-0.55in}
 \includegraphics[angle=180, width=95mm]{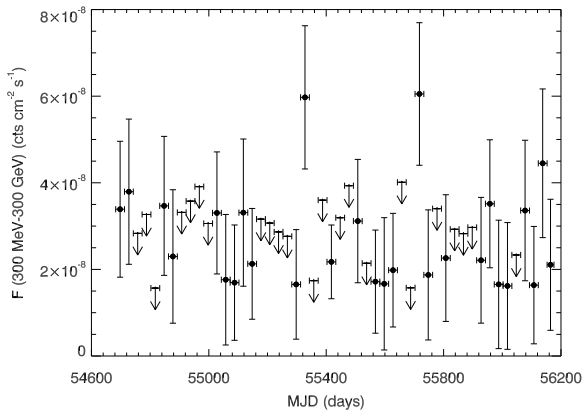}
 \vspace{-0.3in}
 \caption{A binned light curve of 1FGL 1127.7--6244c reveals two significant flaring episodes since the Fermi mission began.  \label{lc}}
 \end{figure}

 \begin{figure}
 \hspace{-0.55in}
 \includegraphics[angle=180, width=95mm]{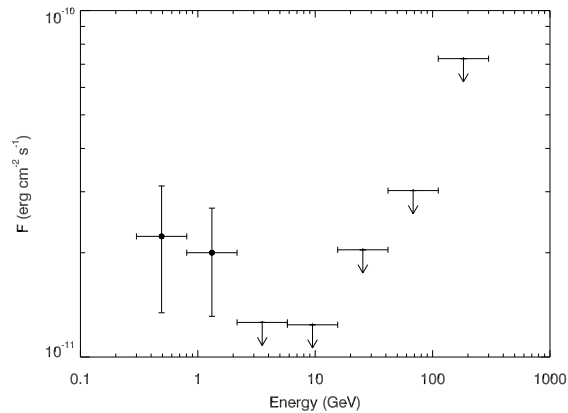}
 \vspace{-0.3in}
 \caption{\textit{Fermi} GeV spectrum of 1FGL 1127.7--6244c during the bright flares.  \label{spectrum}}
 \end{figure}

 \begin{figure}
 \includegraphics[width=75mm]{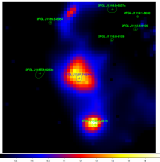}
 \caption{A counts map of 1FGL 1127.7--6244c during the bright flares.  Other nearby sources from the 1FGL and 2FGL catalogs are also labeled.  \label{tsmap}}
 \end{figure}

\section{Observations of 1FGL J1127.7--6244c}

The second correlation between the 1FGL and the \citet{jaschek1982} catalog is the star HD 99771.  Little is known about the Be star other than its spectral type (B7 Vne), and no previously known X-ray sources lie within the Fermi 95\% error ellipse (7.5 arcmin; \citep{nolan2012}.  We observed the field with \textit{XMM-Newton} on 2012 February 4.  The resulting MOS2 image is shown in Figure \ref{mos2}, and we find no keV source coincident with the Be star's position.  


Since the Fermi mission began, 1FGL 1127.7--6244c has experienced two significant flaring episodes.  The 0.3--300 GeV light curve is shown in Figure \ref{lc}.  The GeV spectrum and counts map of the source are shown in Figures \ref{spectrum} and \ref{tsmap}, respectively.  No counterpart to this source was noted in the 2FGL catalog \citep{nolan2012}.

\bigskip 
\begin{acknowledgments}
We gratefully acknowledge support from Fermi Cycle 4 through NASA grant NNX11AO41G and from the National Science Foundation under grant number AST-1109247.  The CTIO 1.5m telescope is operated by the SMARTS Consortium.  
\end{acknowledgments}

\bigskip 

\end{document}